\documentclass[a4paper, 12pt]{article} 
\usepackage[margin=20mm, top=0.5in, bottom=1in]{geometry}
\usepackage{multirow}
\usepackage{array}
\usepackage{graphicx}
\usepackage{authblk}
\usepackage{booktabs}
\usepackage{amsmath}
\usepackage{hyperref}
\usepackage[table,xcdraw]{xcolor}

\providecommand{\keywords}[1]
{
  \small	
  \textbf{\textit{Keywords---}} #1
}

\title{Using Bayesian Evidence Synthesis Methods to Incorporate Real World Evidence in Surrogate Endpoint Evaluation}

\author[1]{Lorna Wheaton}
\author[1,2]{Anastasios Papanikos}
\author[3]{Anne Thomas}
\author[4]{Sylwia Bujkiewicz}

\date{}

\affil[1]{Biostatistics Research Group, Department of Health Sciences, University of Leicester, University Road, Leicester LE1 7RH, UK}

\affil[2]{GlaxoSmithKline R\&D Centre, GlaxoSmithKline, Stevenage, UK}

\affil[3]{Leicester Cancer Research Centre, University of Leicester, Leicester, UK}

\newpage

\begin{document}

\maketitle

\section*{Abstract}
\textbf{Objective:} Traditionally validation of surrogate endpoints has been carried out using RCT data. However, RCT data may be too limited to validate surrogate endpoints. In this paper, we sought to improve validation of surrogate endpoints with the inclusion of real world evidence (RWE). 
\vskip 0in
\noindent 
\textbf{Study Design and Setting:} We use data from comparative RWE (cRWE) and single arm RWE (sRWE), to supplement RCT evidence for evaluation of progression free survival (PFS) as a surrogate endpoint to overall survival (OS) in metastatic colorectal cancer (mCRC). Treatment effect estimates from RCTs, cRWE and matched sRWE, comparing anti-angiogenic treatments with chemotherapy, were used to inform surrogacy patterns and predictions of the treatment effect on OS from the treatment effect on PFS. 

\vskip 0in 
\noindent
\textbf{Results:} Seven RCTs, four cRWE studies and three matched sRWE studies were identified. The addition of RWE to RCTs reduced the uncertainty around the estimates of the parameters for the surrogate relationship. Addition of RWE to RCTs also improved the accuracy and precision of predictions of the treatment effect on OS obtained using data on the observed effect on PFS. 

\vskip 0in
\noindent
\textbf{Conclusion:} The addition of RWE to RCT data improved the precision of the parameters describing the surrogate relationship between treatment effects on PFS and OS and the predicted clinical benefit. 

\keywords{Surrogate Endpoint, Real world evidence, Bayesian meta-analysis, Colorectal cancer}

\newpage

\section{Introduction}

Surrogate endpoints are often used when it takes too long, is too expensive or too difficult to observe the effect on the final clinical outcome of interest~\cite{burzykowski2006}. However, before surrogate endpoints can be used, for example, for regulatory approvals, they should be validated~\cite{dawoud2021raising, bujkiewicz2019nice}. Surrogate endpoints can be validated based on three levels of association; (1) biological plausibility, (2) individual-level surrogacy and (3) trial-level surrogacy~\cite{elston2009use}. However, identifying and validating potential surrogate endpoints can be difficult when data for such analysis are limited. Traditionally, surrogate endpoint evaluation has been carried out using only data from randomised controlled trials (RCTs). Shortages of RCT data are becoming more common as precision medicine evolves and treatments become more effective and are targeted to specific patient populations, leading to smaller cohorts of patients where it takes longer to observe the treatment effect on the final outcome with reasonable precision. This is due to fewer events (such as deaths) recorded in patients receiving targeted therapies and thus high uncertainty around the effectiveness estimates and, as a consequence, around the estimates of association between the treatment effects on surrogate endpoint and final outcome. It is therefore possible that a putative surrogate endpoint cannot be validated and treatments may not be granted conditional approval based on treatment effects on the questionable surrogate endpoint. However, in recent years there has been increased interest in the use of real world evidence (RWE) at all stages of drug development~\cite{beaulieu2020examining,flynn2021marketing,berger2017good,makady2018using,verde2015combining, efthimiou2017combining, welton2020chte2020, schmitz2013incorporating}. This is because RWE can increase the evidence base for decision-making, increase follow-up times and is often more generalisable to the target population. Addition of RWE could improve validation of surrogate endpoints which could not be validated using RCT data alone. 

In this paper, we explored how RWE can be used to strengthen the evidence base for surrogate endpoint evaluation. We made use of comparative RWE (cRWE) and single arm RWE (sRWE) to supplement RCT data on the effectiveness of anti-angiogenic therapies in metastatic colorectal cancer (mCRC). We then investigated the impact of the addition of RWE on the estimates of the surrogate relationship between treatment effects on PFS and OS. 

The remainder of this paper is structured as follows. Data sources and the statistical methods are described in Section 2. The results are presented in Section 3 which is followed by discussion and conclusions in Sections 4 and 5. 

\section{Methods}

\subsection{Data Sources}

\subsubsection{Randomised Controlled Trials}

Data were obtained from a prior literature review conducted by Ciani et al~\cite{ciani2015meta}, which included treatment effect estimates from 11 RCTs in mCRC that assessed anti-angiogenic treatments such as Bevacizumab, combined with various chemotherapies, such as FOLFOX. In the review, Ciani et al~\cite{ciani2015meta}, defined OS as time from randomisation to time of death and PFS as time from randomisation to tumour progression or death from any cause. 

For this paper, we extracted the treatment effects (logHRs) on PFS and OS. RCTs were only included in the analysis where the control arm was chemotherapy and the treatment arm was an anti-angiogenic treatment plus chemotherapy.

\subsubsection{Comparative Real World Evidence}

We carried out a literature review to identify cRWE evaluating anti-angiogenic treatments for mCRC. The following combinations of terms were used to search for studies published between 2000 and 2020 in the PubMed database: (1) ``metastatic colorectal cancer",  (2) ``cohort", ``cohort study", ``retrospective" or ``prospective", (3) ``PFS", (4) ``OS" and (5) ``antiangiogenic" or ``bevacizumab". 

Abstracts, titles and, where necessary, full articles were screened and studies which were not relevant were removed. LogHRs on PFS and OS and their corresponding standard errors were extracted from the remaining studies. To account for potential bias, cRWE studies were only included if they reported treatment effects adjusted for baseline characteristics or potential confounders. 

\subsubsection{Single Arm Real World Evidence}

Papanikos et al~\cite{Papanikos2021} identified 16 single-arm observational studies evaluating anti-angiogenic treatments alone. To obtain relative treatment effects required for surrogate endpoint evaluation, we carried out a literature review of single-arm studies of chemotherapy alone, which subsequently were used as control arms. The following combinations of terms were used to search for single-arm studies on chemotherapy published between 2000 and 2020 in the PubMed database: (1)  ``metastatic colorectal cancer", (2) ``chemotherapy", (3) ``cohort", ``cohort study", ``retrospective" or ``prospective", (4) ``progression" or ``PFS" and (5) ``overall survival" or ``OS". 

The following terms could not be contained in the title or abstract of the studies; ``antiangiogenic", ``bevacizumab", ``cetuximab", ``aflibercept", ``randomised trial", ``randomized trial", or ``phase". These terms were excluded to prevent RCTs and cRWE being returned. Any additional studies found outside the database search were also included.

\subsection{Matching Single Arm Studies}

Unlike RCTs and cRWE, treatment effects cannot be extracted from sRWE as single-arm studies do not make comparisons. To obtain relative treatment effects, sRWE studies were matched using aggregate level data according to the method proposed by Schmitz et al~\cite{schmitz2018use}. The distance $\Delta_{tot}$ between any two single-arm studies $j$ and $k$ was determined as the weighted average of differences in covariates. 

\begin{equation}
\Delta_{t o t}[j, k]=\frac{\sum_{i=1}^{n} w_{i} \cdot \Delta_{i}[j, k]}{\sum_{i=1}^{n} w_{i}}
\end{equation}

\noindent
where $n$ is the number of covariates, $w_{i}$ refers to the weights given to the covariates and $\Delta_{i}[j, k]$ is the normalised difference between studies $j$ and $k$ in covariate $i$. This distance takes a value between $0$ and $1$, where smaller values indicate more similar studies. Distance measures between treatment arms for RCTs and cRWE were also calculated. Since there is no consensus on a threshold for similarity, the maximum distance metric between arms of RCTs was used as the threshold. Where multiple matches were possible, matches with the smallest distance measure were used. The weight of each of the covariates was decided based on rankings from a consensus statement~\cite{goey2018consensus}. 

\subsection{Obtaining Treatment Effects for Matched Single Arm Studies}

\href{https://automeris.io/WebPlotDigitizer/}{WebPlotDigitizer} was used to extract data from Kaplan-Meier curves from each arm of the matched sRWE studies. Kaplan-Meier curves from RCTs and cRWE were also digitized to compare digitized and reported logHRs. Data from risk tables, reporting the number of patients at risk in each arm at regular time intervals, was also extracted to improve approximated IPD~\cite{tierney2007practical}. Where risk tables were not reported, the number of patients and total number of events in each arm were used.  

Data extracted from Kaplan-Meier curves and risk tables (where available) were used in Stata to reconstruct IPD using the \textbf{ipdfc} command by Wei and Royston~\cite{wei2017reconstructing}. The Cox proportional hazards model with a single covariate for treatment arm was used to analyse the reconstructed IPD to obtain logHRs on PFS and OS for matched sRWE studies, cRWE and RCTs.

\subsection{Statistical Analysis}

The standard model for surrogate endpoint evaluation by Daniels and Hughes~\cite{daniels1997meta}, denoted here as D\&H, and bivariate random-effects meta-analysis (BRMA) using the product normal formulation (PNF) were used as alternative methods to model the correlated treatment effects (logHRs) on the surrogate endpoint (PFS) and final outcome (OS) using a Bayesian framework. The models were applied to RCT data alone, RCTs and cRWE and RCTs, cRWE and matched sRWE. Sensitivity analyses to vague prior distributions were conducted for both models. 

\subsubsection{Daniels and Hughes Model}

Daniels and Hughes proposed that the observed treatment effects measured on the surrogate endpoint ($Y_{1 i}$) and final outcome ($Y_{2 i}$) come from a bivariate normal distribution and estimate the underlying true effects on the surrogate and final outcomes ($\delta_{1 i}$ and $\delta_{2 i}$) from each study $i$ with corresponding within-study standard deviations ($\sigma_{1 i}$ and $\sigma_{2 i}$) and within-study correlation ($\rho_{w i}$);

\begin{equation}
\left(\begin{array}{c}
Y_{1 i} \\
Y_{2 i}
\end{array}\right) \sim N\left(\left(\begin{array}{l}
\delta_{1 i} \\
\delta_{2 i}
\end{array}\right),\left(\begin{array}{cc}
\sigma_{1 i}^{2} & \sigma_{1 i} \sigma_{2 i} \rho_{w i} \\
\sigma_{1 i} \sigma_{2 i} \rho_{w i} & \sigma_{2 i}^{2}
\end{array}\right)\right)
\end{equation}

The true effects measured on the surrogate endpoint ($\delta_{1 i}$) are assumed to be independent in each study (fixed effects). It is also assumed there is a linear relationship between the true treatment effects on the final outcome and the surrogate endpoint. 

\begin{equation}
\delta_{2 i} \mid \delta_{1 i} \sim N\left(\lambda_{0}+\lambda_{1} \delta_{1 i}, \psi_{2}^{2}\right)
\end{equation}

Daniels and Hughes referred to a surrogate relationship as perfect when the following conditions were met: (a) $\lambda_{0}=0$, (b) $\lambda_{1}\neq0$ and (c) $\psi_{2}^{2}=0$. These conditions state that (a) no treatment effect on the surrogate endpoint implies no treatment effect on the final outcome, (b) the slope is not zero implying an association between treatment effects on the surrogate and final outcomes and (c) treatment effects on the final outcome can be perfectly predicted by treatment effects on the surrogate endpoint. 

To implement this model in a Bayesian framework, non-informative prior distributions were placed on the fixed effects $\mathrm{\delta}_{1i} \sim \mathrm{N}\left(0,10^{4}\right)$ and regression parameters $\mathrm{\lambda}_{0,1} \sim \mathrm{N}\left(0,10^{4}\right)$. To ensure a non-informative prior distribution on the conditional variance, a uniform prior distribution was placed on the conditional standard deviation $\mathrm{\psi_{2}} \sim \mathrm{Unif}\left(0,2\right)$. A minimally informative prior distribution $\mathrm{\rho_{wi}} \sim \mathrm{Unif}\left(0,1\right)$ was placed on the within-study correlation. 

\subsubsection{Bivariate Random-Effects Meta-Analysis (Product Normal Formulation)}

The D\&H model does not estimate correlation or the study-level $R^2$ which are often used to assess the strength of a surrogate relationship~\cite{buyse2000,burzykowski2001,renfro2012}. For example, the German Institute for Quality and Efficiency in Healthcare (IQWIG) defined an acceptable surrogate endpoint by setting a lower bound for the confidence interval on the correlation coefficient to be 0.85~\cite{iqwig}. To be able to estimate the between-studies correlation and the study-level $R^2$, we used BRMA PNF as an alternative method to model the surrogacy pattern. BRMA PNF has the same within-study model as the D\&H model (2) but the between-studies model assumes exchangeability of the correlated true (random) treatment effects on both outcomes. In the PNF, the bivariate normal distribution is represented as a sequence of univariate conditional distributions (4):

\begin{equation}
\left\{\begin{array}{l}
\delta_{1 i} \sim N\left(\eta_{1}, \psi_{1}^{2}\right) \\
\delta_{2 i} \mid \delta_{1 i} \sim N\left(\eta_{2 i}, \psi_{2}^{2}\right) \\
\eta_{2 i}=\lambda_{0}+\lambda_{1} \delta_{1 i}
\end{array}\right.
\end{equation}

\noindent
where $\delta_{1 i}$ and $\delta_{2 i}$ are the true effects in the population which are correlated, assumed exchangeable and normally distributed. 

The parameters of the BRMA PNF model can be represented in terms of the parameters of the bivariate normal distribution using the following formulae~\cite{bujkiewicz2016bayesian}. 

\begin{equation}
\label{eq:BRMAlink1}
\psi_{1}^{2}=\tau_{1}^{2}, \psi_{2}^{2}=\tau_{2}^{2}-\lambda_{1}^{2} \tau_{1}^{2}, \lambda_{1}=\frac{\tau_{2}}{\tau_{1}} \rho,
\end{equation}

\noindent
where $\tau_1$ and $\tau_2$ are the between-studies heterogeneity parameters and $\rho$ is the between-studies correlation. To implement this model in a Bayesian framework, vague prior distributions were placed on the between-studies parameters $\mathrm{\tau}_{1,2} \sim \mathrm{Unif}\left(0,2\right)$, $\mathrm{\rho} \sim \mathrm{Unif}\left(-1,1\right)$ and the intercept $\mathrm{\lambda_{0}} \sim \mathrm{N}\left(0,10^{4}\right)$ implying prior distributions on $\lambda_{1}$, $\psi_{1}^{2}$ and $\psi_{2}^{2}$ through rearranging the relationships (5). 

In addition to the surrogacy criteria from the D\&H model, a perfect surrogate relationship is defined in the BRMA PNF model when $\rho=\pm 1$~\cite{bujkiewicz2019bivariate}. This implies a perfect association between treatment effects on the surrogate endpoint and final outcome. The study-level $R^2$ in this random-effects model is equal to $\rho^2$~\cite{renfro2012}.

\subsubsection{Bias Adjustment}

To account for systematic differences in treatment effects between RCTs, cRWE and sRWE, the BRMA PNF model was extended to allow for bias adjustment. The between-studies model (4) remains the same for all studies and the within-study model for RCTs remains the same as in (2). However, the within-study models for cRWE and matched sRWE include bias terms, $\alpha_{j}$ and $\beta_{j}$ respectively, for the surrogate and final outcomes ($j=1,2$): 

\begin{equation}
\left(\begin{array}{c}
Y_{1 i} \\
Y_{2 i}
\end{array}\right) \sim N\left(\left(\begin{array}{l}
\delta_{1 i} + \alpha_{1} \\
\delta_{2 i} + \alpha_{2}
\end{array}\right),\left(\begin{array}{cc}
\sigma_{1 i}^{2} & \sigma_{1 i} \sigma_{2 i} \rho_{w i} \\
\sigma_{1 i} \sigma_{2 i} \rho_{w i} & \sigma_{2 i}^{2}
\end{array}\right)\right)
\end{equation}

\begin{equation}
\left(\begin{array}{c}
Y_{1 i} \\
Y_{2 i}
\end{array}\right) \sim N\left(\left(\begin{array}{l}
\delta_{1 i} + \beta_{1} \\
\delta_{2 i} + \beta_{2}
\end{array}\right),\left(\begin{array}{cc}
\sigma_{1 i}^{2} & \sigma_{1 i} \sigma_{2 i} \rho_{w i} \\
\sigma_{1 i} \sigma_{2 i} \rho_{w i} & \sigma_{2 i}^{2}
\end{array}\right)\right)
\end{equation}

\noindent
Additional non-informative prior distributions were placed on the bias terms,  $\mathrm{\alpha}_{1,2} \sim \mathrm{N}\left(0,10^{4}\right)$ and $\mathrm{\beta}_{1,2} \sim \mathrm{N}\left(0,10^{4}\right)$. 

\subsubsection{Cross-validation}

To assess whether addition of RWE improves accuracy or precision of predictions, a ``take-one-out" cross-validation procedure was conducted for the D\&H and BRMA PNF models. For each study $i~(i=1,...,N)$, the treatment effect on the final outcome, $Y_{2i}$, was removed and assumed missing at random. The treatment effect on the final outcome was predicted from the treatment effect on the surrogate endpoint, conditional on data on both outcomes from all other studies in the meta-analysis. The mean predicted effect is equal to the mean predicted true effect from MCMC simulation. The variance of the predicted effect is $\sigma_{2 i}^{2}+\operatorname{var}\left(\hat{\delta}_{2 i} \mid Y_{1 i}, \sigma_{1 i}, Y_{1(-i)}, Y_{2(-i)}\right)$ where $Y_{1,2(-i)}$ are the observed treatment effects on the surrogate and final outcomes for the remaining studies not omitted in the $ith$ iteration~\cite{daniels1997meta,bujkiewicz2019nice}. For a valid surrogate, the 95\% predicted interval (constructed using the variance) will contain the observed treatment effect in at least 95\% of studies. 

\subsection{Software and Computing}

All models were implemented using WinBUGS~\cite{lunn2000winbugs} where estimates were obtained using MCMC simulation with 150000 iterations (including 50000 burn-in). Convergence was checked via visual assessment of history, density and autocorrelation plots. Posterior estimates are presented as means (approximately normal posterior) or medians (skewed posterior) with 95\% credible intervals (CrI). R was used for data manipulation, to execute WinBUGS code using the R2WinBUGS package~\cite{sturtz2005r2winbugs} and to produce figures using the ggplot2 package. 

\section{Results}

\subsection{Summary of Data}

Of the 11 RCTs obtained from the prior literature review, 4 were excluded for not investigating the effect of anti-angiogenics in combination with chemotherapy against chemotherapy alone. Overall 7 RCTs were included in the analysis.

The database search of PubMed returned 166 publications for cRWE studies and 145 publications for sRWE studies on the chemotherapy arm. After screening titles, abstracts and, where appropriate, full articles, 7 cRWE studies comparing bevacizumab against chemotherapy remained and 8 sRWE studies of chemotherapy alone remained. Of the 7 cRWE studies, 4 adjusted for covariates. 

Five covariates were reported in all sRWE studies and of these 5, sex was the only covariate not recommended for reporting in the consensus statement. Following the consensus statement ranking, sex was given a weight of 1 and all other covariates a weight of 2. The covariates selected for matching were: treatment line (weight=2, current treatment line normalised between range 1-3), age (weight=2, median age normalised between range 18-100), performance score (weight=2, ECOG/WHO score normalised between range 0-3), tumour location (weight=2, proportion with colon tumour compared to rectum tumour) and sex (weight=1, proportion of females).

Table 1 shows the distance measures between the sRWE studies. A maximum distance measure of 0.035 was applied as this was close to the maximum distance measure from RCTs (0.032). This resulted in an exploration of 9\% (n=11) of possible matches. In Table 1, possible matches are shaded and final matches (lowest distance measures) are shown in bold. Overall, three matched sRWE studies were included in the analysis.

\begin{table}
\centering
\setlength{\extrarowheight}{0pt}
\addtolength{\extrarowheight}{\aboverulesep}
\addtolength{\extrarowheight}{\belowrulesep}
\setlength{\aboverulesep}{0pt}
\setlength{\belowrulesep}{0pt}
\caption{Distance Metric between Single Arm Observational Studies}
\label{table:matching}
\resizebox{\linewidth}{!}{%
\begin{tabular}{lcccccccc} 
\toprule
                    & \begin{tabular}[c]{@{}c@{}}Dong \\2015\end{tabular} & \begin{tabular}[c]{@{}c@{}}Matsumoto \\2007\end{tabular} & \begin{tabular}[c]{@{}c@{}}Catalano \\2009\end{tabular} & \begin{tabular}[c]{@{}c@{}}Fuse \\2008\end{tabular} & \begin{tabular}[c]{@{}c@{}}Suenaga \\2008\end{tabular} & \begin{tabular}[c]{@{}c@{}}Fuse \\2007\end{tabular} & \begin{tabular}[c]{@{}c@{}}Hochster \\2003\end{tabular} & \begin{tabular}[c]{@{}c@{}}Yoshino \\2007\end{tabular}  \\ 
\hline
Bendell 2012 (1)    & {\cellcolor[rgb]{0.753,0.753,0.753}}\textbf{0.029}  & 0.273                                                    & 0.048                                                   & 0.134                                               & 0.077                                                  & 0.126                                               & 0.043                                                   & 0.057                                                   \\
Bendell 2012 (2)    & {\cellcolor[rgb]{0.753,0.753,0.753}}0.034           & 0.273                                                    & 0.043                                                   & 0.139                                               & 0.074                                                  & 0.131                                               & 0.041                                                   & 0.062                                                   \\
Hurwitz 2014        & 0.151                                               & 0.156                                                    & 0.158                                                   & 0.211                                               & 0.171                                                  & 0.248                                               & 0.169                                                   & 0.179                                                   \\
Van Cutsem 2009 (1) & 0.051                                               & 0.234                                                    & 0.092                                                   & 0.093                                               & 0.040                                                  & 0.085                                               & 0.073                                                   & {\cellcolor[rgb]{0.753,0.753,0.753}}0.018               \\
Van Cutsem 2009 (2) & {\cellcolor[rgb]{0.753,0.753,0.753}}0.034           & 0.232                                                    & 0.085                                                   & 0.082                                               & 0.044                                                  & 0.074                                               & 0.083                                                   & {\cellcolor[rgb]{0.753,0.753,0.753}}0.018               \\
Van Cutsem 2009 (3) & 0.047                                               & 0.221                                                    & 0.096                                                   & 0.072                                               & 0.045                                                  & 0.064                                               & 0.094                                                   & {\cellcolor[rgb]{0.753,0.753,0.753}}\textbf{0.013}      \\
Van Cutsem 2009 (4) & 0.051                                               & 0.224                                                    & 0.095                                                   & 0.079                                               & 0.045                                                  & 0.071                                               & 0.087                                                   & {\cellcolor[rgb]{0.753,0.753,0.753}}0.017               \\
Bennouna 2017 (1)   & 0.058                                               & 0.227                                                    & 0.069                                                   & 0.110                                               & 0.073                                                  & 0.102                                               & 0.089                                                   & 0.041                                                   \\
Bennouna 2017 (2)   & 0.208                                               & 0.115                                                    & 0.208                                                   & 0.175                                               & 0.204                                                  & 0.212                                               & 0.242                                                   & 0.191                                                   \\
Buchler 2014 (1)    & 0.045                                               & 0.233                                                    & 0.079                                                   & 0.094                                               & 0.070                                                  & 0.086                                               & 0.080                                                   & {\cellcolor[rgb]{0.753,0.753,0.753}}0.025               \\
Buchler 2014 (2)    & 0.052                                               & 0.213                                                    & 0.088                                                   & 0.082                                               & 0.050                                                  & 0.072                                               & 0.091                                                   & {\cellcolor[rgb]{0.753,0.753,0.753}}0.014               \\
Ocvirk 2011 (1)     & 0.043                                               & 0.260                                                    & 0.072                                                   & 0.100                                               & 0.066                                                  & 0.094                                               & 0.066                                                   & 0.065                                                   \\
Ocvirk 2011 (2)     & 0.051                                               & 0.226                                                    & 0.097                                                   & 0.080                                               & 0.047                                                  & 0.062                                               & 0.096                                                   & 0.044                                                   \\
Moriwaki 2012 (1)   & 0.204                                               & 0.062                                                    & 0.233                                                   & 0.121                                               & 0.160                                                  & 0.148                                               & 0.243                                                   & 0.165                                                   \\
Moriwaki 2012 (2)   & 0.184                                               & 0.099                                                    & 0.228                                                   & 0.132                                               & 0.129                                                  & 0.169                                               & 0.211                                                   & 0.150                                                   \\
Kotaka 2016         & 0.068                                               & 0.200                                                    & 0.095                                                   & 0.072                                               & {\cellcolor[rgb]{0.753,0.753,0.753}}\textbf{0.033}     & 0.064                                               & 0.104                                                   & {\cellcolor[rgb]{0.753,0.753,0.753}}0.034               \\
\bottomrule
\end{tabular}
}
\end{table}

Figure 1 presents the data from the included studies as a scatter plot of the observed treatment effects on PFS and OS. The plot shows a possible strong positive relationship between treatment effects on PFS and OS, suggesting a potential valid surrogate relationship.

\begin{figure}
  \includegraphics[width=\linewidth]{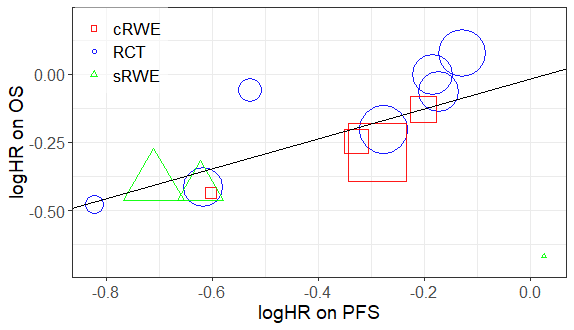}
  \caption{Scatterplot of observed logHR on OS against observed logHR on PFS for RCTs (blue circle), cRWE (red square) and matched sRWE (green triangle). Black line shows the linear relationship between logHR on PFS and logHR on OS obtained from D\&H model conducted using all sources of evidence.}
  \label{fig:bubbleplot}
\end{figure}

\begin{figure*}
  \includegraphics[width=\linewidth]{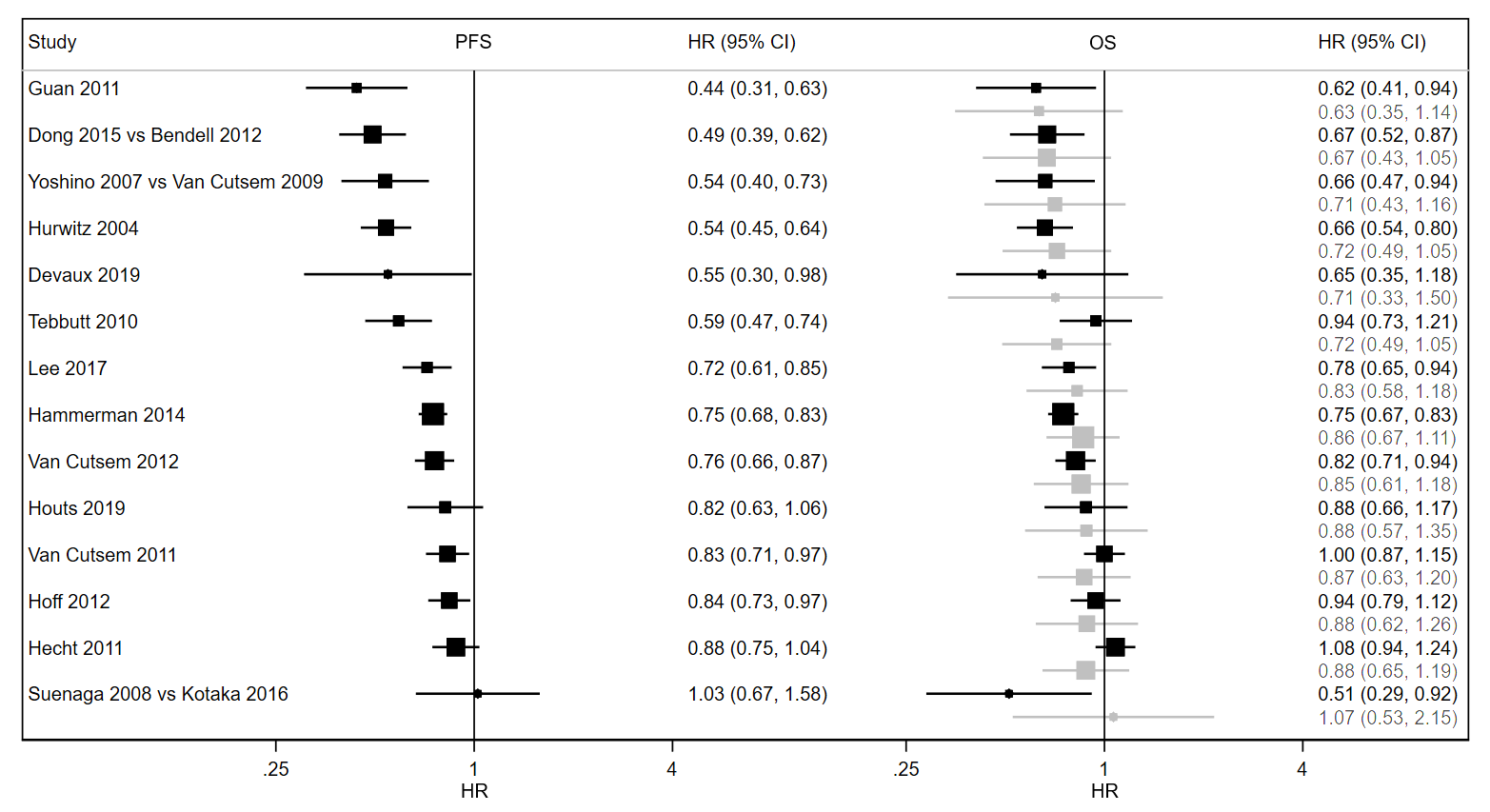}
  \caption{Forest plots of HRs from RCTs, cRWE and sRWE. Graph shows observed HRs and 95\% confidence intervals on PFS (black lines - left) and OS (black lines - right) and predicted HRs and 95\% predicted intervals on OS obtained from cross-validation using D\&H model (grey lines - right).}
  \label{fig:forestplot}
\end{figure*}

\subsection{Daniels and Hughes Model}

Table 2 shows results from the D\&H model. There is evidence of a surrogate relationship regardless of the type of evidence used as the intercept and conditional variance are close to zero and the slope is non-zero. Figure 2 illustrates this relationship, highlighting that studies with larger treatment effects on PFS generally have larger effects on OS.

\begin{table}
\centering
\setlength{\extrarowheight}{0pt}
\addtolength{\extrarowheight}{\aboverulesep}
\addtolength{\extrarowheight}{\belowrulesep}
\setlength{\aboverulesep}{0pt}
\setlength{\belowrulesep}{0pt}
\caption{Surrogacy criteria obtained from D\&H model applied to data from RCTs, comparative RWE (cRWE) and matched single-arm RWE (sRWE). Last two rows give cross-validation results from D\&H model where absolute discrepancy is the absolute difference between the observed logHR and the predicted logHR on OS, while $w_{\hat{Y}_{2j}}/w_{Y_{2j}}$ is the ratio of the width of the 95\% predicted interval of the logHR on OS to the width of the 95\% confidence interval of the observed estimate of the logHR on OS.}
\label{table:D&H}
\resizebox{\linewidth}{!}{%
\begin{tabular}{rrrr} 
\toprule
                                                                                & RCTs                  & RCTs \&  cRWE           & RCTs, cRWE \&  sRWE      \\ 
\hline
\rowcolor[rgb]{0.753,0.753,0.753} $\lambda_0$                                   & 0.10 (-0.13, 0.34)    & 0.051 (-0.13, 0.23)   & -0.015 (-0.20, 0.15)   \\
$\lambda_1$                                                                     & 0.71 (0.13, 1.30)     & 0.69 (0.20, 1.17)     & 0.55 (0.099, 0.97)     \\
\rowcolor[rgb]{0.753,0.753,0.753} $\psi_{2}^{2}$                                & 0.0091 (0.0001, 0.11) & 0.010 (0.0002, 0.051) & 0.010 (0.0002, 0.054)  \\
Absolute Discrepancy, Median (Range)                                            & 0.091 (0.0022, 0.27)  & 0.070 (0.0079, 0.30)  & 0.076 (0.0019, 0.73)   \\
\rowcolor[rgb]{0.753,0.753,0.753} $w_{\hat{Y}_{2j}}/w_{Y_{2j}}$, Median (Range) & 2.90 (1.76, 3.58)     & 2.09 (1.33, 2.40)     & 1.79 (1.22, 2.42)      \\
\bottomrule
\end{tabular}
}
\end{table}

Addition of cRWE to RCTs improved the precision of all three estimates for the surrogate relationship while having minimal impact on the point estimates. For example, using RCT data alone resulted in the conditional variance 0.0091 (95\% CrI: 0.0001, 0.11), addition of cRWE gave a conditional variance of 0.010 (95\% CrI: 0.0002, 0.051). Thus, addition of cRWE reduced uncertainty by 54\% in terms of the credible interval width. Addition of sRWE to RCTs and cRWE further improved precision of the intercept and slope estimates and only slightly decreased precision of the conditional variance. 

\subsection{Bivariate Random-Effects Meta-Analysis (Product Normal Formulation)}

Table 3 shows there was weaker evidence for a surrogate relationship using the BRMA PNF compared to the D\&H model. Although correlation was 0.75 (when using all evidence) the credible interval was wide (95\% CrI: -0.081, 0.98). Furthermore, the credible interval for the slope contained zero, suggesting no relationship between treatment effects on the surrogate and final endpoint. Such differences in results between the models could be a result of the random effects assumption of the BRMA PNF model. When assuming normal random effects is appropriate, greater borrowing of information can lead to more precise estimates. However, when this assumption is violated, the model can lead to over-shrinkage of the true effects, thus potentially reducing the between-studies correlation~\cite{bujkiewicz2017uncertainty}. This can lead to bias and reduce precision of estimates for surrogacy parameters. 

Despite differing results to the D\&H model, addition of RWE generally improved the precision of estimates obtained from the BRMA PNF model. The correlation obtained using RCT data only was 0.75 (95\% CrI: -0.24, 0.99) whereas correlation obtained using all sources of evidence was 0.75 (95\% CrI: -0.081, 0.98). Thus, addition of RWE reduced uncertainty by 14\% in terms of the width of the credible interval while the point estimate remained the same. 

\begin{table}
\centering
\setlength{\extrarowheight}{0pt}
\addtolength{\extrarowheight}{\aboverulesep}
\addtolength{\extrarowheight}{\belowrulesep}
\setlength{\aboverulesep}{0pt}
\setlength{\belowrulesep}{0pt}
\caption{Surrogacy criteria obtained from BRMA PNF model applied to data from RCTs, comparative RWE (cRWE) and matched single-arm RWE (sRWE). Last two rows give cross-validation results from BRMA PNF model where absolute discrepancy is the absolute difference between the observed logHR and the predicted logHR on OS, while $w_{\hat{Y}_{2j}}/w_{Y_{2j}}$ is the ratio of the width of the 95\% predicted interval of the logHR on OS to the width of the 95\% confidence interval of the observed estimate of the logHR on OS.}
\label{table:BRMA}
\resizebox{\linewidth}{!}{%
\begin{tabular}{rrrrr} 
\toprule
                                                                                                                                                              & \begin{tabular}[c]{@{}r@{}}\\RCTs\end{tabular} & \begin{tabular}[c]{@{}r@{}}\\RCTs \& cRWE\end{tabular} & \begin{tabular}[c]{@{}r@{}}\\RCTs, cRWE \& sRWE\end{tabular} & \begin{tabular}[c]{@{}r@{}}Bias Adjusted\\RCTs, cRWE \& sRWE\end{tabular}  \\
\rowcolor[rgb]{0.753,0.753,0.753} $d_1$                                                                                                                       & -0.36 (-0.62, -0.13)                           & -0.34 (-0.49, -0.21)                                & -0.37 (-0.51, -0.23)                                      & -0.36 (-0.57, -0.16)                                                    \\
$d_2$                                                                                                                                                         & -0.14 (-0.35, 0.046)                           & -0.18 (-0.31, -0.052)                               & -0.22 (-0.34, -0.11)                                      & -0.14 (-0.29, -0.0020)                                                  \\
\rowcolor[rgb]{0.753,0.753,0.753} $\rho$                                                                                                                      & 0.75 (-0.24, 0.99)                             & 0.66 (-0.30, 0.97)                                  & 0.75 (-0.081, 0.98)                                       & 0.69 (-0.32, 0.99)                                                      \\
$\tau_1$                                                                                                                                                      & 0.27 (0.11, 0.59)                              & 0.19 (0.079, 0.36)                                  & 0.22 (0.11, 0.38)                                         & 0.23 (0.10, 0.43)                                                       \\
\rowcolor[rgb]{0.753,0.753,0.753} $\tau_2$                                                                                                                    & 0.21 (0.075, 0.48)                             & 0.16 (0.072, 0.31)                                  & 0.17 (0.086, 0.30)                                        & 0.15 (0.048, 0.28)                                                      \\
$\lambda_0$                                                                                                                                                   & 0.054 (-0.21, 0.33)                            & 0.0068 (-0.26, 0.28)                                & -0.012 (-0.25, 0.22)                                      & 0.011 (-0.22, 0.26)                                                     \\
\rowcolor[rgb]{0.753,0.753,0.753} $\lambda_1$                                                                                                                 & 0.54 (-0.16, 1.31)                             & 0.54 (-0.25, 1.40)                                  & 0.56 (-0.053, 1.19)                                       & 0.42 (-0.17, 1.14)                                                      \\
$\psi_{2}^{2}$                                                                                                                                                & 0.013 (0.0009, 0.10)                           & 0.011 (0.0014, 0.048)                               & 0.011 (0.0010, 0.050)                                     & 0.0077 (0.0003, 0.047)                                                  \\
\rowcolor[rgb]{0.753,0.753,0.753} $R^{2}$                                                                                                                     & 0.57 (0.0045, 0.98)                            & 0.43 (0.0023, 0.94)                                 & 0.56 (0.0061, 0.97)                                       & 0.49 (0.0022, 0.97)                                                     \\
$\alpha_1$                                                                                                                                                    & -                                              & -                                                   & -                                                         & 0.041 (-0.32, 0.38)                                                     \\
\rowcolor[rgb]{0.753,0.753,0.753} $\alpha_2$                                                                                                                  & -                                              & -                                                   & -                                                         & -0.12 (-0.36, 0.12)                                                     \\
$\beta_1$                                                                                                                                                     & -                                              & -                                                   & -                                                         & -0.13 (-0.52, 0.29)                                                     \\
\rowcolor[rgb]{0.753,0.753,0.753} $\beta_2$                                                                                                                   & -                                              & -                                                   & -                                                         & -0.28 (-0.59, 0.039)                                                    \\
\begin{tabular}[c]{@{}r@{}}Absolute Discrepancy,\\Median (Range)\end{tabular}                                                                                 & 0.16 (0.017, 0.26)                             & 0.16 (0.0046, 0.23)                                 & 0.12 (0.026, 0.57)                                        & 0.17 (0.0092, 0.66)                                                     \\
\rowcolor[rgb]{0.753,0.753,0.753} \begin{tabular}[c]{@{}>{\cellcolor[rgb]{0.753,0.753,0.753}}r@{}}$w_{\hat{Y}_{2j}}/w_{Y_{2j}}$,\\Median (Range)\end{tabular} & 2.74 (1.61, 3.30)                              & 2.09 (1.15, 2.49)                                   & 1.78 (1.14, 2.56)                                         & 1.66 (1.14, 3.07)                                                       \\
\bottomrule
\end{tabular}
}
\end{table}

\subsection{Bias Adjustment}

Table 3 shows the results of the bias adjusted BRMA PNF model. When adjusting for bias, there was no improvement in precision and the 95\% credible interval for the slope estimate still contained zero, indicating weak evidence for surrogacy. In addition, when using bias adjustment, correlation was lower (0.69; 95\% CrI: -0.32, 0.99) than when not using bias adjustment (0.75; 95\% CrI: -0.081, 0.98), providing less evidence for a valid surrogate relationship. The lower correlation could be explained by the bias adjustment reducing between-studies heterogeneity for treatment effects on the final outcome.

\subsection{Cross-validation}

The last two rows of Table 2 show the results of cross-validation using the D\&H model. Median absolute discrepancy between predicted and observed treatment effects on OS decreased with the addition of cRWE and slightly increased with the further addition of sRWE. This increase is likely explained by matched Suenaga 2008 and Kotaka 2016 studies which had a very low relative treatment effect on OS in comparison to the treatment effect on PFS (Figures 1 and 2). Table 2 also shows that the length of the predicted interval, relative to the observed confidence interval, fell with the addition of cRWE (2.90 to 2.09) and sRWE (2.09 to 1.79). Overall, the precision of the predicted intervals improved with the addition of RWE. Cross-validation for the BRMA PNF model showed similar results (Table 3). 

\section{Discussion}

When existing clinical trial data are limited, surrogate endpoint validation may fail. As a result, new therapies may not receive conditional marketing authorisation or, if approved, they may still fail at the health technology assessment (HTA) decision-making stage, by HTA agencies such as the National Institute for Health and Care Excellence (NICE)~\cite{dawoud2021raising,bujkiewicz2019nice}. In this paper, we provide an approach for using RWE to strengthen the evidence base for surrogate endpoint validation. 

There are several limitations of this research. Inclusion of sRWE relied on digitizing Kaplan-Meier curves. However, such curves are not always published and thus, potentially useful studies would not be included. One method to overcome this issue is to extract median survival times for the surrogate and final outcomes and use an exponential hazard assumption, as proposed by Schmitz et al~\cite{schmitz2018use}, to obtain treatment effects. In a preliminary analysis we applied this method to the mCRC dataset; however, the exponential hazard assumption did not provide a good approximation. 

Matching of sRWE was based on study level covariates and thus was prone to bias as patients were not randomised or matched at the individual-level and, therefore, the assumption of exchangeability may have been violated~\cite{leahy2019incorporating}. This bias was further exacerbated by only matching on 5 covariates, when 10 additional characteristics were recommended by the consensus statement and identified as risk factors. However, these variables were not reported in the included studies. 

While the bias adjustment was used to account for potential systematic differences in treatment effects between data sources, all sources of evidence contributed the same weight to the model. This suggests that RCTs, cRWE and matched sRWE were of equivalent quality. However, it is widely acknowledged that RCTs are above RWE in the hierarchy of evidence~\cite{brighton2003hierarchy}. Further methodological research is carried out to allow for accounting for such differences in quality.

\section{Conclusions}

RWE can be used to improve the precision of estimates for surrogate endpoint validation relative to using RCT data alone. The addition of RWE to RCT data also allows for more precise predictions to be made of the treatment effects on the final clinical outcome based on the treatment effect measured on the surrogate endpoint.

\section*{Acknowledgements}
LW was funded by a National Institute for Health Research (NIHR) Pre-doctoral Fellowship [NIHR301013]. The views expressed are those of the authors and not necessarily those of the NIHR or the Department of Health and Social Care. The funding bodies played no role in the design of the study and collection, analysis, and interpretation of data and in writing the manuscript.
SB was funded by Medical Research Council [grant no. MR/R025223/1].

\section*{Author Contributions}
SB conceptualised the study and provided supervision to LW and TP. LW and TP curated the data. LW carried out all formal analyses, methodological development and visualisation, and provided the original draft of the manuscript. AT provided clinical advice. All authors contributed to the manuscript revisions.

\bibliographystyle{unsrt}
\bibliography{references.bib}

\end{document}